\documentclass[twocolumn,showpacs,prc,superscriptaddress,floatfix,amsmath,amssymb]{revtex4-1}
\usepackage{graphicx}
\usepackage{dcolumn}
\usepackage{bm}
\usepackage[usenames]{color}
\draft

\begin{document}

\title{Search for the $\Theta^{+}$ pentaquark via the $\pi^-p\rightarrow K^-X$ reaction at 1.92 GeV/$c$}

\author{K.~Shirotori}
\affiliation{Department of Physics, Tohoku University, Sendai 980-8578, Japan}
\altaffiliation[Corresponding author, email: sirotori@post.j-parc.jp, Present address: ]{Advanced Science Research Center (ASRC), Japan Atomic Energy Agency (JAEA), Tokai, Ibaraki 319-1195, Japan}

\author{T.~N.~Takahashi}
\affiliation{Department of Physics, University of Tokyo, 7-3-1 Hongo, Tokyo 113-0033, Japan}

\author{S.~Adachi}
\affiliation{Department of Physics, Kyoto University, Kyoto 606-8502, Japan}

\author{M.~Agnello}
\affiliation{Dipartimento di Fisica, Politecnico di Torino, I-10129 Torino, Italy}

\author{S.~Ajimura}
\affiliation{Research Center for Nuclear Physics (RCNP), 10-1 Mihogaoka, Ibaraki, Osaka 567-0047, Japan}

\author{K.~Aoki}
\affiliation{Institute of Particle and Nuclear Studies (IPNS), High Energy Accelerator Research Organization (KEK), Tsukuba, 305-0801, Japan}

\author{H.~C.~Bang}
\affiliation{Department of Physics and Astronomy, Seoul National University, Seoul 151-747, Korea}

\author{B.~Bassalleck}
\affiliation{Department of Physics and Astronomy, University of New Mexico, New Mexico 87131-0001, USA}

\author{E.~Botta}
\affiliation{Dipartimento di Fisica, Universit$\grave{a}$ di Torino, I-10125 Torino, Italy}

\author{S.~Bufalino}
\affiliation{Dipartimento di Fisica, Universit$\grave{a}$ di Torino, I-10125 Torino, Italy}

\author{N.~Chiga}
\affiliation{Department of Physics, Tohoku University, Sendai 980-8578, Japan}

\author{P.~Evtoukhovitch}
\affiliation{Joint Institute for Nuclear Research, Dubna, Moscow Region 141980, Russia}

\author{A.~Feliciello}
\affiliation{INFN, Istituto Nazionale di Fisica Nucleare, Sez. di Torino, I-10125 Torino, Italy}

\author{H.~Fujioka}
\affiliation{Department of Physics, Kyoto University, Kyoto 606-8502, Japan}

\author{F.~Hiruma}
\affiliation{Department of Physics, Tohoku University, Sendai 980-8578, Japan}

\author{R.~Honda}
\affiliation{Department of Physics, Tohoku University, Sendai 980-8578, Japan}

\author{K.~Hosomi}
\affiliation{Department of Physics, Tohoku University, Sendai 980-8578, Japan}

\author{Y.~Ichikawa}
\affiliation{Department of Physics, Kyoto University, Kyoto 606-8502, Japan}

\author{M.~Ieiri}
\affiliation{Institute of Particle and Nuclear Studies (IPNS), High Energy Accelerator Research Organization (KEK), Tsukuba, 305-0801, Japan}

\author{Y.~Igarashi}
\affiliation{Institute of Particle and Nuclear Studies (IPNS), High Energy Accelerator Research Organization (KEK), Tsukuba, 305-0801, Japan}

\author{K.~Imai}
\affiliation{Advanced Science Research Center (ASRC), Japan Atomic Energy Agency (JAEA), Tokai, Ibaraki 319-1195, Japan}

\author{N.~Ishibashi}
\affiliation{Department of Physics, Osaka University, Toyonaka 560-0043, Japan}

\author{S.~Ishimoto}
\affiliation{Institute of Particle and Nuclear Studies (IPNS), High Energy Accelerator Research Organization (KEK), Tsukuba, 305-0801, Japan}

\author{K.~Itahashi}
\affiliation{RIKEN, 2-1 Hirosawa, Wako, Saitama 351-0198, Japan}

\author{R.~Iwasaki}
\affiliation{Institute of Particle and Nuclear Studies (IPNS), High Energy Accelerator Research Organization (KEK), Tsukuba, 305-0801, Japan}

\author{C.~W.~Joo}
\affiliation{Department of Physics and Astronomy, Seoul National University, Seoul 151-747, Korea}

\author{M.~J.~Kim}
\affiliation{Department of Physics and Astronomy, Seoul National University, Seoul 151-747, Korea}

\author{S.~J.~Kim}
\affiliation{Department of Physics and Astronomy, Seoul National University, Seoul 151-747, Korea}

\author{R.~Kiuchi}
\affiliation{Department of Physics and Astronomy, Seoul National University, Seoul 151-747, Korea}

\author{T.~Koike}
\affiliation{Department of Physics, Tohoku University, Sendai 980-8578, Japan}

\author{Y.~Komatsu}
\affiliation{Department of Physics, University of Tokyo, 7-3-1 Hongo, Tokyo 113-0033, Japan}

\author{V.~V.~Kulikov}
\affiliation{ITEP, Institute of Theoretical and Experimental Physics, Moscow 117218, Russia}

\author{S.~Marcello}
\affiliation{Dipartimento di Fisica, Universit$\grave{a}$ di Torino, I-10125 Torino, Italy}

\author{S.~Masumoto}
\affiliation{Department of Physics, University of Tokyo, 7-3-1 Hongo, Tokyo 113-0033, Japan}

\author{K.~Matsuoka}
\affiliation{Department of Physics, Osaka University, Toyonaka 560-0043, Japan}

\author{K.~Miwa}
\affiliation{Department of Physics, Tohoku University, Sendai 980-8578, Japan}

\author{M.~Moritsu}
\affiliation{Department of Physics, Kyoto University, Kyoto 606-8502, Japan}

\author{T.~Nagae}
\affiliation{Department of Physics, Kyoto University, Kyoto 606-8502, Japan}


\author{M.~Naruki}
\affiliation{Institute of Particle and Nuclear Studies (IPNS), High Energy Accelerator Research Organization (KEK), Tsukuba, 305-0801, Japan}

\author{M.~Niiyama}
\affiliation{Department of Physics, Kyoto University, Kyoto 606-8502, Japan}

\author{H.~Noumi}
\affiliation{Research Center for Nuclear Physics (RCNP), 10-1 Mihogaoka, Ibaraki, Osaka, 567-0047, Japan}

\author{K.~Ozawa}
\affiliation{Institute of Particle and Nuclear Studies (IPNS), High Energy Accelerator Research Organization (KEK), Tsukuba, 305-0801, Japan}

\author{N.~Saito}
\affiliation{Institute of Particle and Nuclear Studies (IPNS), High Energy Accelerator Research Organization (KEK), Tsukuba, 305-0801, Japan}

\author{A.~Sakaguchi}
\affiliation{Department of Physics, Osaka University, Toyonaka 560-0043, Japan}

\author{H.~Sako}
\affiliation{Advanced Science Research Center (ASRC), Japan Atomic Energy Agency (JAEA), Tokai, Ibaraki 319-1195, Japan}

\author{V.~Samoilov}
\affiliation{Joint Institute for Nuclear Research, Dubna, Moscow Region 141980, Russia}

\author{M.~Sato}
\affiliation{Department of Physics, Tohoku University, Sendai 980-8578, Japan}

\author{S.~Sato}
\affiliation{Advanced Science Research Center (ASRC), Japan Atomic Energy Agency (JAEA), Tokai, Ibaraki 319-1195, Japan}

\author{Y.~Sato}
\affiliation{Institute of Particle and Nuclear Studies (IPNS), High Energy Accelerator Research Organization (KEK), Tsukuba, 305-0801, Japan}

\author{S.~Sawada}
\affiliation{Institute of Particle and Nuclear Studies (IPNS), High Energy Accelerator Research Organization (KEK), Tsukuba, 305-0801, Japan}

\author{M.~Sekimoto}
\affiliation{Institute of Particle and Nuclear Studies (IPNS), High Energy Accelerator Research Organization (KEK), Tsukuba, 305-0801, Japan}

\author{H.~Sugimura}
\affiliation{Department of Physics, Kyoto University, Kyoto 606-8502, Japan}

\author{S.~Suzuki}
\affiliation{Institute of Particle and Nuclear Studies (IPNS), High Energy Accelerator Research Organization (KEK), Tsukuba, 305-0801, Japan}

\author{H.~Takahashi}
\affiliation{Institute of Particle and Nuclear Studies (IPNS), High Energy Accelerator Research Organization (KEK), Tsukuba, 305-0801, Japan}

\author{T.~Takahashi}
\affiliation{Institute of Particle and Nuclear Studies (IPNS), High Energy Accelerator Research Organization (KEK), Tsukuba, 305-0801, Japan}

\author{H.~Tamura}
\affiliation{Department of Physics, Tohoku University, Sendai 980-8578, Japan}

\author{T.~Tanaka}
\affiliation{Department of Physics, Osaka University, Toyonaka 560-0043, Japan}

\author{K.~Tanida}
\affiliation{Department of Physics and Astronomy, Seoul National University, Seoul 151-747, Korea}
\affiliation{Advanced Science Research Center (ASRC), Japan Atomic Energy Agency (JAEA), Tokai, Ibaraki 319-1195, Japan}

\author{A.~O.~Tokiyasu}
\affiliation{Department of Physics, Kyoto University, Kyoto 606-8502, Japan}

\author{N.~Tomida}
\affiliation{Department of Physics, Kyoto University, Kyoto 606-8502, Japan}

\author{Z.~Tsamalaidze}
\affiliation{Joint Institute for Nuclear Research, Dubna, Moscow Region 141980, Russia}

\author{M.~Ukai}
\affiliation{Department of Physics, Tohoku University, Sendai 980-8578, Japan}

\author{K.~Yagi}
\affiliation{Department of Physics, Tohoku University, Sendai 980-8578, Japan}

\author{T.~O.~Yamamoto}
\affiliation{Department of Physics, Tohoku University, Sendai 980-8578, Japan}

\author{S.~B.~Yang}
\affiliation{Department of Physics and Astronomy, Seoul National University, Seoul 151-747, Korea}

\author{Y.~Yonemoto}
\affiliation{Department of Physics, Tohoku University, Sendai 980-8578, Japan}

\author{C.~J.~Yoon}
\affiliation{Department of Physics and Astronomy, Seoul National University, Seoul 151-747, Korea}

\author{K.~Yoshida}
\affiliation{Department of Physics, Osaka University, Toyonaka 560-0043, Japan}

\begin{abstract}
The $\Theta^+$ pentaquark baryon was searched for via the $\pi^-p\rightarrow K^-X$ reaction 
in a missing-mass resolution of 1.4 MeV/$c^2$(FWHM) at J-PARC. 
$\pi^-$ meson beams were incident on the liquid hydrogen target with the beam momentum of 1.92 GeV/$c$. 
No peak structure corresponding to the $\Theta^+$ mass was observed. 
The upper limit of the production cross section averaged over the scattering angle of 2$^{\circ}$ to 15$^{\circ}$ 
in the laboratory frame was obtained to be 0.26 $\mu$b/sr in the mass region of 1.51$-$1.55 GeV/$c^2$.
The upper limit of the $\Theta^+$ decay width using the effective Lagrangian approach 
was obtained to be 0.72 MeV/$c^2$ and 3.1 MeV/$c^2$ for $J^P_{\Theta}=1/2^+$ and $J^P_{\Theta}=1/2^-$, respectively. 
\end{abstract}

\maketitle

The Laser-Electron Photon facility at SPring-8 (LEPS) collaboration reported the first evidence of the $\Theta^+$ pentaquark\cite{Nakano}. 
A peak was observed in the missing mass spectrum of the $\gamma{n}\rightarrow K^-X$ reaction on a $^{12}$C target at a mass of $1.54\pm0.01$ GeV/$c^2$. 
This baryon resonance should have an exotic quark content of $uudd\bar{s}$. 
The observed peak width was consistent with the experimental resolution of 25 MeV/$c^2$, 
which suggests the intrinsic width is quite narrow. 
On the theory side, D.~Diakonov~${\it et\;al.}$ predicted a baryon resonance 
with a narrow width of less than 15 MeV/$c^2$ at 1.53 GeV/$c^2$ using a chiral soliton model\cite{Diakonov}. 
Good agreement between this theoretical prediction and the LEPS group's result triggered investigations of $\Theta^+$ by research groups world-wide. 
The LEPS group's first evidence was immediately confirmed and re-examined experimentally by research groups at various facilities\cite{ReportsE}. 
For example, the DIANA collaboration reported an evidence for $\Theta^+$ in the $K^+{\mathrm{Xe}}\rightarrow K^0{p}{X}$ reaction with a significance of 4.4$\sigma$\cite{DIANA1}. 
The CLAS collaboration reported a peak with a significance of 7.8$\sigma$ using the $\gamma{p}\rightarrow \pi^+K^-K^+n$ reaction\cite{CLASp}. 
The LEPS group's second data showed a 5.1$\sigma$ peak from the $\gamma{d}\rightarrow K^-K^+np$ reaction in the energy range of 2.0$-$2.4 GeV. 
The statistics was improved by a factor of 8 compared to the first measurement\cite{Nakano2}. 
The DIANA collaboration also enlarged the statistics of the peak with the mass resolution of 8.2 MeV/$c^2$(FWHM). 
From the production cross section, the intrinsic decay width of $\Gamma_{\Theta \rightarrow NK}=0.36\pm0.11$ MeV/$c^2$\cite{DIANA2} was estimated. 
Experiments such as these that reported positive results had a statistical significance ranging between 3 and 8; 
however, at the same time, many negative results with high statistics also been reported from high-energy experiments.
For example, the BABAR collaboration reported that the upper limits of the $\Theta^+$ yield per $q\overline{q}$ event was obtained to be $5.0{\times}10^{-5}$ at $\sqrt{s}$ = 10.58 GeV, 
which was a factor of 8 below the typical yield for ordinary baryons\cite{BABAR}. 
Therefore, the existence of $\Theta^+$ has remained quite controversial. 
It should be also noted that there are no experiments which measured the intrinsic width of $\Theta^+$ directly. 
Therefore, it has been long awaited for other experiments having much higher statistics 
and a high sensitivity to measure the intrinsic width in different reactions and different experimental setups 
in order to have a definite answer to the existence of $\Theta^+$.

The structure of $\Theta^+$ has been also studied intensively in various theoretical models\cite{ReportsT}. 
The observed mass and the remarkably narrow width should be explained well. 
There is an idea\cite{Jaffe} to explain the narrow width by introducing a correlation among the constituent quarks. 
However, none of the existing theoretical models have reasonably succeeded so far. 
It is an important challenge to our knowledge about the quark dynamics 
at low energy understanding the decay mechanism and structure of $\Theta^+$.


A cutting-edge approach to investigating $\Theta^+$ is the use of hadron-induced reactions. 
Such reactions have a large production cross section, 
and their production mechanism is more straightforward compared to that of photo-induced reactions. 
In this light, experiments on the $\Theta^+$ production using a hadron beam such as $\pi^{-}$ and $K^{+}$ have attracted considerable attention.
In fact, searches for $\Theta^+$ using hadron beams have already been performed 
at the High Energy Accelerator Research Organization (KEK) 12 GeV Proton Synchrotron 
via the $\pi^-{p}\rightarrow K^- X$~\cite{Miwa1} and the $K^+{p}\rightarrow \pi^+ X$ reactions\cite{Miwa2}. 
Unfortunately, these experiments failed to obtain clear evidence of $\Theta^+$.
However, a peak structure at 1.53 GeV/$c^2$ with 2.6$\sigma$ significance was reported in the $\pi^-{p}\rightarrow K^- X$ reaction at $p_{\pi}=1.92$ GeV/$c$. 
The obtained upper limit of the differential cross section was 2.9 $\mu$b/sr at the 90\% confidence level 
assuming that $\Theta^+$ is produced isotropically in the center-of-mass system. 
The 90\% C.L. upper limit of the cross section of the $K^+{p}\rightarrow \pi^+ X$ reaction using a 1.20 GeV/$c$ $K^+$ beam was obtained 
to be 3.5 $\mu$b/sr for the mass region of 1.51$-$1.55 GeV/$c^2$. 
It implies that the contribution of the $K^*$ exchange term was small. 
This result was consistent with that obtained for the $\gamma{p}\rightarrow \bar{K}^0K^+n$ reaction by the CLAS collaboration\cite{CLAS3}. 
From those results, the nucleon pole term that is sensitive to the decay width of $\Theta^+$ was considered to be dominant in the production cross section\cite{Oh}. 
A narrow width implied that the production cross section was expected to be much smaller than that in usual hadron production. 
Therefore, an experiment with higher sensitivity is required to confirm the existence of $\Theta^+$.


The Japan Proton Accelerator Research Complex (J-PARC) E19\cite{Naruki} experiment aims to 
to search for the $\Theta^+$ pentaquark with high-resolution and high-statistics via the $\pi^-p\rightarrow K^-X$ reaction. 
This was the first experiment to be carried out at the K1.8 beam line in the Hadron Facility\cite{HDHall}. 
By using a high-intensity pion beam and a high-resolution spectrometer system at this beam line, 
a sensitivity of 75 nb/sr in the laboratory frame could be achieved for a narrow $\Theta^+$ ($\Gamma_{\Theta}<$ 2 MeV/$c^2$). 
In this paper, we report on the results obtained from the first set of experiments, that were carried out in 2010. 
The beam momentum was set to 1.92 GeV/$c$ in order to re-examine the previous experimental result. 
The secondary pion beam intensity per pulse was typically 1.0$\times10^{6}$ for the present experiment. 
In total, 7.8$\times10^{10}$ $\pi^-$ mesons were incident on the liquid hydrogen target with a mass thickness of 0.86 g/cm$^2$.

\begin{figure}[t]
\begin{center}
\includegraphics[width=210pt]{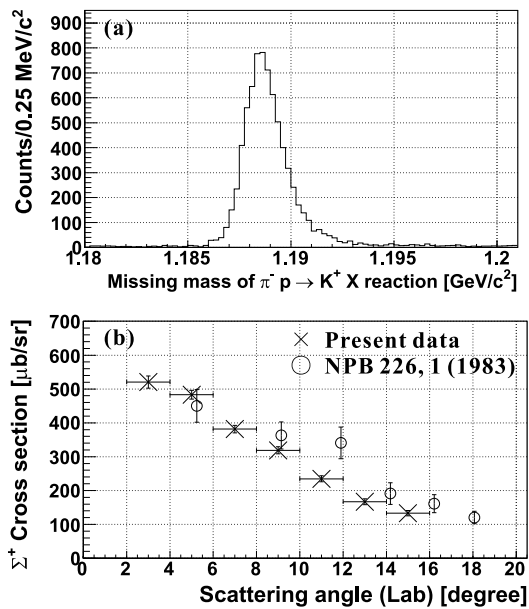}
\caption[]{
(a): The missing mass spectrum of the $\pi^{+}{p}\rightarrow K^+ X$ reactions at 1.37 GeV/$c$. 
The $\Sigma^+$ peak is clearly observed. 
(b): The differential cross sections of the $\Sigma^+$ production. 
The present data and the angular distribution reported in the old experiment\cite{CSSigmaP} are indicated by crosses 
and open circles, respectively. 
}
\label{Sigma}
\end{center}
\end{figure}

The incident pions were identified and momentum-analyzed using the K1.8 beam line spectrometer, 
which has a momentum resolution of less than 0.1\%(FWHM). 
The beam pions were separated by using the time-of-flight 
between timing counters placed at both the entrance and the exit of the beam line spectrometer. 
The electrons remaining in the beam were rejected by using a gas $\check{\mathrm{C}}$erenkov counter ($n=1.002$) placed 
at the most upstream point in the beam line spectrometer. 
The amount of electron contamination in the pion beam was less than 0.1\%. 
The muons contaminating in the beam was estimated to be 3.5\% from the Decay Turtle code\cite{TURTLE}. 
The beam momentum was reconstructed by two sets of beam line chambers placed at the entrance and the exit of the beam line spectrometer.

The scattered kaons were identified and momentum-analyzed by the Superconducting Kaon Spectrometer (SKS) system 
with an angular acceptance of 100 msr and a momentum resolution of 0.2\%(FWHM). 
The scattered particles were identified by using an aerogel $\check{\mathrm{C}}$erenkov counter ($n=1.05$) 
and an acrylic $\check{\mathrm{C}}$erenkov counter ($n=1.49$) at the trigger level. 
The precise identification was carried out in the offline analysis 
by using the time-of-flight technique in combination with information about the flight path 
and the reconstructed momentum obtained through the SKS system. 
The momentum was also calculated from data obtained from the two sets of chambers 
placed at the entrance and the exit of the SKS magnet. 
The SKS magnetic field was set at 2.5 T, 
and scattered particles with a momentum of 0.7$-$1.0 GeV/$c$ 
and scattering angle was measured from $2^{\circ}$ to $15^{\circ}$ in this system. 
The very forward angle was not used due to the worse vertex resolution. 
The reaction vertex point was extracted from the closest distance between the tracks of beam pions and scattered kaons. 
The remaining background events due to other target cell materials was estimated to be $2.8\pm0.1$\%.


\begin{figure} [t]
\includegraphics[width=230pt]{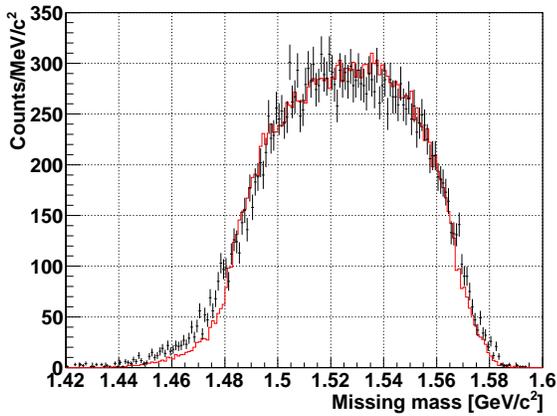}
\caption[]{
The missing mass spectrum and the background shape 
for the $\pi^- p \rightarrow K^-X$ reaction at the beam momentum of 1.92 GeV/$c$. 
The black points with error bars are the experimental data. 
The contribution of the simulated background is indicated by red histogram.
}
\label{MissMass}
\end{figure}

To evaluate various parameters of the spectrometer system, 
such as the missing mass resolution, absolute mass scale, detection efficiencies, and kaon survival rate, 
the known $\Sigma^{\pm}$ production was also measured via the $\pi^{\pm}{p}\rightarrow K^+\Sigma^{\pm}$ reactions at 1.37 GeV/$c$ 
in order to cover the same momentum region of scattered kaons from the $\pi^-p\rightarrow K^-\Theta^+$ reaction at 1.92 GeV/$c$. 
Figure~\ref{Sigma}(a) shows the missing mass spectrum of the $\pi^{+}{p}\rightarrow K^+X$ reaction showing a clear peak of $\Sigma^+$.
The missing mass resolution for $\Sigma^{+}$ was $1.9\pm0.1$ MeV/$c^{2}$(FWHM), 
which corresponds to a resolution of $1.4\pm0.1$ MeV/$c^{2}$(FWHM) for $\Theta^+$. 
The energy loss of both the beam and the scattered particles in the target was corrected based on a simulation using the Bethe-Bloch formula. 
From the $\Sigma^{\pm}$ data and by measuring the beam which passed through both spectrometer system, 
the uncertainty for the absolute mass scale is estimated to be $\pm$1.7 MeV/$c^2$, including that of the energy loss correction of $\pm$0.3 MeV/$c^2$. 
The cross section was estimated from the yield of $\Sigma^{\pm}$ taking all the experimental efficiencies and the kaon survival rate into account. 
The cross section of the $\Sigma^{+}$ production obtained in this experiment is consistent with the old experimental data\cite{CSSigmaP}, as shown in Fig~\ref{Sigma}(b).


\begin{figure} [t]
\includegraphics[width=230pt]{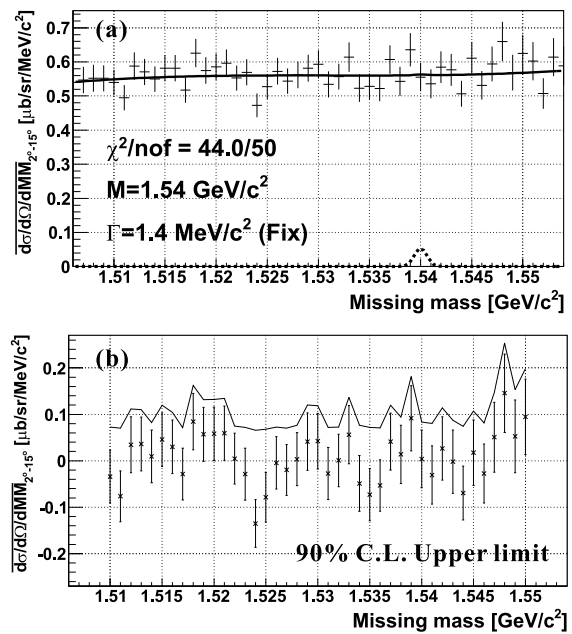}
\caption[]{(a): The differential cross section as a function of missing mass of the $\pi^- p \rightarrow K^- \Theta^+$ reaction. 
The fitted result with a Gaussian and a third order polynomial background shape is indicated by the solid line. 
In the fit, a Gaussian peak shape function whose peak is fixed at 1.54 GeV/$c^2$ 
and a background shape of a third-order polynomial function were used. 
The width was fixed to be the experimental resolution of 1.4 MeV/$c^2$ (FWHM). 
The peak with a 90\% confidence level is also shown in the dotted line. 
(b): The differential cross section of the $\pi^- p \rightarrow K^-\Theta^+$ reaction averaged over 2$^{\circ}$ to 15$^{\circ}$ in the laboratory frame 
with the $\Theta^+$ width fixed at 1.4 MeV/$c^2$ (FWHM). 
The black line indicates the upper limit of the differential cross section at 90\% confidence level. 
For the calculation of the line position, the amplitude for the Gaussian peak is constrained to be a positive value. 
The systematic error is included in the error bars. 
}
\label{UpperLimit}
\end{figure}

The missing mass spectrum for the $\pi^- p \rightarrow K^- X$ reaction is shown in Fig.~\ref{MissMass}. 
No structure corresponding to $\Theta^+$ was observed in the spectrum. 
The spectrum was fitted with a Gaussian and a third-order polynomial background shape 
to obtain the upper limit of the total cross section of the $\Theta^+$ production as a function of the missing mass. 
Figure~\ref{UpperLimit}(a) shows the differential cross section 
for the $\pi^- p \rightarrow K^- \Theta^+$ reaction averaged over 2$^{\circ}$ to 15$^{\circ}$ in the laboratory frame. 
In the fitting, a fixed missing mass resolution of 1.4 MeV/$c^2$(FWHM) was assumed in the mass region of 1.51$-$1.55 GeV/$c^2$. 
In the figure, a peak at 1.54 GeV/$c^2$ with 90\% confidence level is also shown. 
The upper limit of the differential cross section averaged over 2$^{\circ}$ to 15$^{\circ}$ in the laboratory frame 
was obtained to be 0.26 $\mu$b/sr at 90\% confidence level in the region of 1.51$-$1.55 GeV/$c^2$, as shown in Fig.~\ref{UpperLimit}(b). 
Compared with the previous KEK experimental result of 2.9 $\mu$b/sr, more than 10 times higher sensitivity has been achieved. 
The systematic error of the differential cross section is 8$-$11\%, 
which originates from the stability of the detector efficiencies, 
simulations of kaon survival rate and kaon absorption by detector materials, and acceptance correction. 
The major contribution to errors in the mass region of 1.54$-$1.55 GeV/$c^2$ is the acceptance correction 
for the scattering angle around 15$^{\circ}$, which is near the edge of the SKS acceptance. 
The typical value of the correction error is 8\%. 
For other mass region, the contribution of the correction error is smaller than that of the other systematic errors, which is estimated to be 6\%. 
The associated background shape is reproduced by the Monte Carlo simulation using the Geant4 code\cite{Geant4}, 
including the acceptance of the SKS system, decay of scattered $K^-$, distribution of the beam profile, and beam momentum bite. 
The main backgrounds in the $\pi^- p \rightarrow K^-X$ reaction are the $\phi$ production, $\Lambda$(1520) production, 
and $K^-K^+n$ and $K^-K^0p$ productions following the three-body phase space. 
The cross section of each process obtained from the old experiments was adjusted within the errors (20$-$30\%) to reproduce the data shape. 
The simulation result agrees well with the backgrounds of the present data.


\begin{table}[t]
\caption{
The total cross section calculated by the pseudo-scalar (PS) and pseudo-vector (PV) scheme 
using the static ($Fs$) and covariant ($Fc$) type form factors\cite{Hyodo2}. 
The incident energy for the $\pi$- and $K$-induced reaction is 1.92 GeV/$c$ and 1.20 GeV/c, respectively. 
The cross section was calculated by integrating over the whole solid angle. 
The decay width of $\Theta^+$ was set to be 1 MeV/$c^2$ at a mass of 1.54 GeV/$c^2$. 
}
\begin{center}
\begin{tabular}{c|cc|cc} \hline \hline
$J^P_{\Theta}=1/2^+$ & $\pi$ induced & & $K$ induced & \\
                     & PS [$\mu$b] & PV [$\mu$b] & PS [$\mu$b] & PV [$\mu$b] \\ \hline
$Fs$ $\Lambda_s=0.5$ GeV  & 9.2   & 0.51 & 119 & 9.6 \\ 
$Fc$ $\Lambda_c=1.8$ GeV  & 5.3   & 0.29 & 595 & 46  \\ \hline \hline
$J^P_{\Theta}=1/2^-$ & $\pi$ induced & & $K$ induced & \\
                     & PS [$\mu$b] & PV [$\mu$b] & PS [$\mu$b] & PV [$\mu$b] \\ \hline
$Fs$ $\Lambda_s=0.5$ GeV  & 0.18   & 0.40 & 1.9 & 4.2 \\ 
$Fc$ $\Lambda_c=1.8$ GeV  & 0.10   & 0.23 & 9.6 & 20  \\ \hline 
\end{tabular}
\label{table1}
\end{center}
\end{table}

The cross section of the $\Theta^+$ production was theoretically calculated by hadronic models using effective Lagrangians and form factors. 
In these models, two schemes were used to introduce the Yukawa coupling, namely, the pseudo-scalar (PS) and pseudo-vector (PV) scheme. 
The coupling constants and the form factors were used as free parameters whose values are based on the known hyperon reactions. 
To include the finite size effect of a hadron, 
static ($F_{s}=\frac{\Lambda_s^2}{\Lambda_s^2+q^2}$) 
and covariant [$F_{c}=\frac{\Lambda_c^4}{\Lambda_c^4+(x-m_x)^4}$] type form factors were introduced in the calculation. 
The $\Theta^+$ production mechanism was comprehensively studied via the $\gamma N, NN, KN$, and $\pi N$ reactions near the production threshold\cite{Oh}, \cite{Ko}. 
The cross section of the $\pi^-{p}\rightarrow K^-\Theta^+$ reaction was calculated to be 9 $\mu$b 
using the PS scheme and the static form factor at a pion beam momentum of 1.92 GeV/$c$, 
assuming the width of $\Theta^+$ to be 1 MeV/$c^2$ and $J^P_{\Theta}=1/2^+$. 
The production cross section of $\Theta^+$ via both the $\pi^-{p}\rightarrow K^-\Theta^+$ and $K^+{p}\rightarrow \pi^+\Theta^+$ reactions 
was systematically studied in \cite{Hyodo1}, \cite{Hyodo2}. 
Since the $K^*$ exchange term is considered to be extremely small based on experimental results\cite{Miwa2}, \cite{CLAS3} 
and only the nucleon pole term contributes to the reactions, 
the total production cross section is directly proportional to the decay width of $\Theta^+$. 
The total production cross sections integrated over the whole solid angle were calculated as listed in Table~\ref{table1}, 
assuming a decay width of 1 MeV/$c^2$ at 1.54 GeV/$c^2$. 
This calculation explained the previous experimental results obtained at KEK\cite{Miwa1}, \cite{Miwa2}.

The upper limit of the total cross section was calculated from the measured forward-angle differential cross section of the present experimental result 
taking into account the spectrometer acceptance and predicted angular distributions. 
We adopted the meson-induced production model\cite{Hyodo2} to determine the upper limit of the decay width of $\Theta^+$. 
Because this model predicts three different angular distributions depending on the coupling scheme and $J^P$ of $\Theta^+$, 
the respective upper limits of the total cross section are calculated to be 0.31 $\mu$b ($1/2^+$ PS and $1/2^-$ PS), 0.21 $\mu$b ($1/2^+$ PV), and 0.26 $\mu$b ($1/2^-$ PV). 
Compared with the total cross sections theoretically calculated with a $\Theta^+$ width of 1 MeV/$c^2$, 
the upper limits of the $\Theta^+$ decay width were respectively found to be 0.72 MeV/$c^2$ and 3.1 MeV/$c^2$ 
for $J^P_{\Theta}=1/2^+$ and $J^P_{\Theta}=1/2^-$.

From the experimental results of both the current $\pi$-induced and the previous $K$-induced reactions, 
the upper limit of the decay width was theoretically calculated.
In the case of $J^P_{\Theta}=1/2^+$, the model in the PV scheme using the static form factor ($F_s$) is favored to explain the small upper limit obtained in both reactions. 
In this case, the upper limit of the total cross section of the $\pi^-{p}\rightarrow K^-\Theta^+$ reaction
was estimated to be 0.21 $\mu$b using the angular distribution of the model. 
Compared with the theoretically predicted values in Table~\ref{table1}, 
the limit of the width was extracted to be 0.41 MeV/$c^2$. 
In the case of $J^P_{\Theta}=1/2^-$, the PS scheme using the static form factor ($F_s$) is favored. 
In this case, the upper limit of the production cross section for the $1/2^-$ state is estimated to be 0.31 $\mu$b. 
This corresponds to an upper limit of the decay width of 1.7 MeV/$c^2$.


In summary, the pentaquark $\Theta^+$ was searched for via the $\pi^-p\rightarrow K^-X$ reaction at the K1.8 beam line in the J-PARC Hadron Facility. 
A missing mass resolution of $1.4\pm0.1$ MeV/$c^2$ for the $\Theta^+$ mass of 1.54 GeV/$c^2$ was achieved. 
In the missing mass spectrum, no peak structure corresponding to $\Theta^{+}$ mass was observed. 
The upper limit of the differential cross section averaged over the scattering angle of 2$^{\circ}$ to 15$^{\circ}$ in the laboratory frame 
was less than 0.26 $\mu$b/sr at the 90\% confidence level in the mass region of 1.51$-$1.55 GeV/$c^2$. 
Combining the theoretical calculation with the experimental result obtained from the $\pi^-{p}\rightarrow K^-\Theta^+$ reactions, 
the upper limit of the decay width was respectively calculated to be 0.72 MeV/$c^2$ and 3.1 MeV/$c^2$ 
for $J^P_{\Theta}=1/2^+$ and $J^P_{\Theta}=1/2^-$.


We would like to acknowledge the outstanding efforts of 
the staff of the J-PARC accelerator and the Hadron experimental facility without whom this experiment would not be possible. 
We also acknowledge T. Hyodo for the helpful theoretical discussions. 
Two of the authors (K.S. and T.N.T) thank the Japan Society for the Promotion of Science (JSPS) for their support. 
This work was supported by a Grant-in-Aid for Scientific Research on Priority Areas (No.~17070005) 
and a Grant-in-Aid for Scientific Research on Innovative Areas (No.~22105512) 
from the Ministry of Education, Culture, Sports, Science and Technology, Japan. 
We acknowledge support from National Research Foundation and WCU program of the Ministry Education Science and Technology (Korea). 



\end{document}